\pdfoutput=1

\documentclass[11pt]{article}

\usepackage[]{acl}
\usepackage{acl}

\usepackage{times}
\usepackage{graphicx}
\usepackage{amsmath}
\usepackage{amssymb}
\usepackage{booktabs}
\usepackage{multirow}
\usepackage{latexsym}
\usepackage{enumitem}
\usepackage[T1]{fontenc}
\usepackage[utf8]{inputenc}
\usepackage{microtype}
\usepackage[ruled,linesnumbered]{algorithm2e}
\usepackage{algpseudocode}
\usepackage{hyperref}

\allowdisplaybreaks     %允许多行公式分页

\title{CACL: Community-Aware Heterogeneous Graph Contrastive Learning for Social Media Bot Detection}

\author{Sirry Chen$^\spadesuit$\thanks{\quad Equal contribution.}\hspace{0.5em}, Shuo Feng$^\spadesuit$$^{*}$, Songsong Liang$^\spadesuit$, Chen-Chen Zong$^\spadesuit$, Jing Li$^\heartsuit$, Piji Li$^\spadesuit$\thanks{\quad Corresponding author.}\hspace{0.5em} \\
        $^\spadesuit$Nanjing University of Aeronautics and Astronautics \\
        $^\heartsuit$The Hong Kong Polytechnic University\\
        \texttt{\{sirrychen,pjli\}@nuaa.edu.cn}}

\begin{document}
\maketitle
\begin{abstract}
Social media bot detection is increasingly crucial with the rise of social media platforms. Existing methods predominantly construct social networks as graph and utilize graph neural networks (GNNs) for bot detection. However, most of these methods focus on how to improve the performance of GNNs while neglecting the community structure within social networks. Moreover, GNNs based methods still face problems such as poor model generalization due to the relatively small scale of the dataset and over-smoothness caused by information propagation mechanism. To address these problems, we propose a \textbf{C}ommunity-\textbf{A}ware Heterogeneous Graph \textbf{C}ontrastive \textbf{L}earning framework (\textbf{CACL}), which constructs social network as heterogeneous graph with multiple node types and edge types, and then utilizes community-aware module to dynamically mine both hard positive samples and hard negative samples for supervised graph contrastive learning with adaptive graph enhancement algorithms. Extensive experiments demonstrate that our framework addresses the previously mentioned challenges and outperforms competitive baselines on three social media bot benchmarks.
\end{abstract}

\section{Introduction}
Social media bots, widely found within social networks, are software systems designed to interact with humans~\cite{orabi2020bot_defination}, e.g., 9\% to 15\% active users on Twitter exhibit bot-like behaviors~\cite{varol2017bot_exist}. Although social media bots have brought convenience to people's daily life, including news broadcasts, online customer service, etc., their negative applications, such as online disinformation~\cite{cresci2023bot_misinformation}, election interference~\cite{cresci2020bot_election1,ferrara2017bot_election2}, extremism advocacy~\cite{berger2015extreme_action}, seem to be more widespread and absolutely cannot be overlooked. To solve the above issues, numerous methods have been proposed to detect social media bots efficiently.

\begin{figure}[t]
    \centering
    \includegraphics[width=1\linewidth]{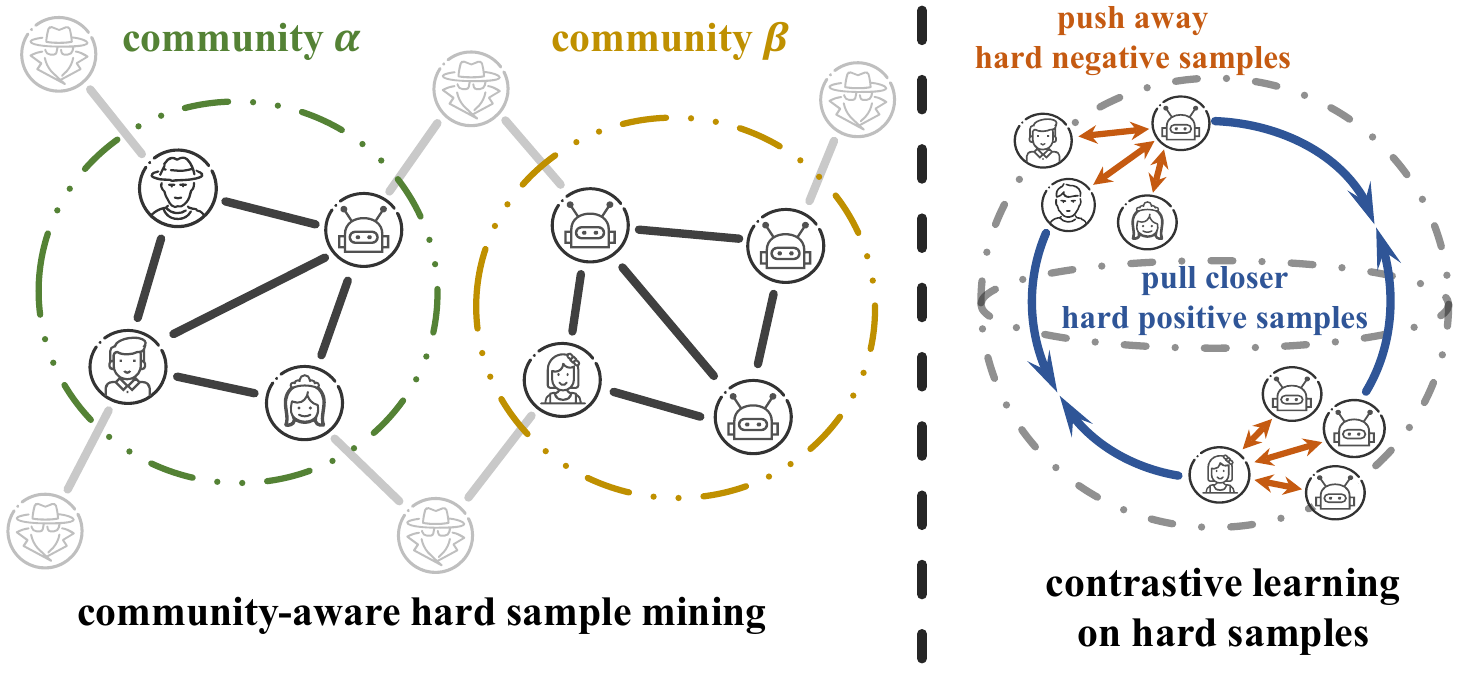}
    \caption{Illustration of considering diff-class nodes within one community as hard negative samples while same-class nodes between communities as hard positive samples. Based on these, graph contrastive learning is employed to push hard negative samples away and pull hard positive samples closer on the hypersphere.}
    \label{figure:small figure of idea}
\end{figure}

Existing works for social media bot detection primarily consist of feature based and graph based methods. In recent years, because feature engineering~\cite{yang2020scalable_feature_engi,feng2021satar,dehghan2023node_structural} has been proven to be incapable of addressing the challenge of generalization~\cite{feng2022twibot-22}, graph convolutional neural networks (GCNs)~\cite{kipf2016GCN} based models have been designed to leverage graph structure information. While homogeneous graphs~\cite{ali2019detect_me_if_you_can_gcn} fail to model diverse relationships between social accounts, such as following and commenting, methods based on heterogeneous graphs, e.g., heterogeneous GCNs~\cite{zhao2020heterogeneous_gcn}, relational GCNs~\cite{feng2021botrgcn}, relational graph transformer~\cite{feng2022RGT}, etc., can better extract node features by capturing information from various types of relationships.
% both intra-relation and inter-relation information.

However, most of these efforts primarily focus on improving the performance of node feature extraction by designing sophisticated GNNs, while neglecting the inherent characteristic of social networks, i.e., social networks naturally divide into groups of users with denser connections inside each group and fewer connections between different groups~\cite{nguyen2014CommImportant,cherifi2019community_structure_in_graph,liu2023botmoe}.
% ~\citet{newman2004newman_CommDense} and ~\citet{tu2018user_same_interest} further highlight that users within one social community usually share common interests, consequently, they tend to interact more frequently with each other than with others outside their community. 
For example, under a politically related topic, users tend to interact more frequently with each other than with others who are not interested in this topic~\cite{newman2004newman_CommDense}. Intuitively, these users form a political community. And in this community, to influence public opinion or achieve other objectives, bots often share political opinions and interact with humans. As a result, they exhibit shared interests and characteristics with humans in the same community, while bots from other communities are relatively different from the humans in this community~\cite{tu2018user_same_interest}. This phenomenon makes bots from other communities can be easily distinguished from users in this political community, once we identify the community structures.
% \pjli{so we should consider what more information?}

% The graph structure information mentioned above is important to understand the patterns of social media~\cite{comm_structure_important} and can be efficiently extracted by community-aware methods through partitioning vast social network into small-scale subgraphs~\cite{rossetti2018graph_structure_by_comm}. However, it is worth noting that humans and bots inevitably coexist in every community.
% % Whereas, the crucial point is that both humans and bots are inevitably present in every community.
% For example, under a politically relevant tweet, it is often observed that responses from bots are intermingled with those from humans~\cite{cresci2020bot_election1}. 
While leveraging these community structures can facilitate bot detection, it also introduces problems. When we treat users as nodes, due to the over-smoothing problem~\cite{chen2020GCN_smooth}, GNNs based methods will result in highly similar representations for nodes in the same community, and different representations for nodes in differernt communities. Therefore, diff-class nodes in one community tend to have short distances in the feature space, while same-class nodes in different communities tend to have long distances between them~\cite{zhou2020GCN_comm_smooth}. As illustrated in Figure~\ref{figure:small figure of idea}, we define diff-class nodes in one community as hard negative samples and same-class nodes in different communities as hard positive samples. These hard samples bring difficulties to bot detection.

To address the above challenges, we propose the \textbf{C}ommunity-\textbf{A}ware Heterogeneous Graph \textbf{C}ontrastive \textbf{L}earning framework (i.e., \textbf{CACL})\footnote{\href{https://github.com/SirryChen/CACL}{https://github.com/SirryChen/CACL}}, a graph contrastive learning framework capable of perceiving community structures. In general, this framework is divided into graph level and node level components. Firstly, we design a community-aware module to perform community detection and mine both hard negative and positive samples. Then, supervised graph contrastive learning is introduced to handle these hard samples, by guiding GNNs model to focus on subtle differences among hard negative samples and push them away, and reasonably disregard the distinct features among hard positive samples and pull them closer. Besides, the underlying feature extraction networks in community-aware module are shared with the networks in contrastive learning module, allowing us to perform dynamic and adaptive hard sample mining. In this manner, by continuously updating model parameters and mining new hard samples, the representations of same-class nodes are consistently gathered on the hypersphere while those of diff-class are separated, which is beneficial to classification. To construct contrastive graphs, we introduce link prediction, synonymy substitution, and node feature shifting that are adaptive to graph topology structure, text information, and users' attributes, respectively. We summarize our main contributions as follows:
\begin{itemize}[leftmargin=*]
  \item We design a community-aware module to dynamically extract community structures within social networks, and perform both hard negative and positive sample mining based on these structures.
  \item We propose a graph contrastive learning framework with community structure-aware capacity, and jointly perform data augmentation on topology, text and attribute levels to generate contrastive graphs.
  \item Extensive experiments on three social media bot detection datasets  demonstrate our framework can significantly improve the performance of several GNNs backbones, which bears out the effectiveness and superiority of \textbf{CACL} framework.
\end{itemize}

\section{Related work}
\subsection{Graph-based Bot Detection}
Traditional neural networks~\cite{kudugunta2018deep,wu2021deep} struggle to directly handle non-Euclidean structured data while GNNs provide an effective way. \citet{ali2019detect_me_if_you_can_gcn} conceptualize users as nodes and construct a graph with single type of relationships between users. \citet{feng2022RGT} utilize the Relation Graph Transformer which employs a semantic attention mechanism to aggregate information from different types of edges. \citet{yang2023ROSGAS} construct a heterogeneous graph with diverse relationships between these users and their texts. Despite their success, existing methods still face problems such as poor model generalization and over-smoothness~\cite{ma2021GCN_generalization}. To mitigate the above issues, our approach leverages the community-aware module to mine both hard negative and positive samples and employs graph contrastive learning to deal with these samples.

\subsection{Community Detection on Social Networks}
Early methods, such as Louvain algorithm~\cite{blondel2008louvain}, utilize only graph structure information and aim to maximize modularity during community detection. In recent years, deep neural networks have been introduced to combine node features with graph structure information. 
Methods such as Markov Random Fields~\cite{he2021Community-centri}, Graph Autoencoders (GAE)~\cite{wang2020GAE_cd}, and Variational Graph Autoencoders (VGAE)~\cite{salha2022modularityCD} are employed to capture graph structure information and conduct unsupervised community detection.
% ~\citet{he2021Community-centri} incorporate Markov Random Fields and GCNs. ~\citet{salha2022modularityCD} integrate modularity information into Graph Autoencoders (GAE) and Variational Graph Autoencoders (VGAE). 
Different from~\citet{tan2023botpercent}, who split whole social networks into several vast communities, we follow the original definition of community and consider each community as a subgraph.

\subsection{Graph Contrastive Learning}
Contrastive learning originates in the field of image processing and has been widely applied to non-Euclidean graph processing in recent years. GRACE~\cite{zhu2020GRACE} employs feature mask and edge removal to generate augmented views and regards the same node in two views as a positive pair. CBD~\cite{zhou2023detect_on_fly} adopts contrastive learning in the pre-training stage to extract knowledge from unlabeled data and uses a consistency loss in the fine-tuning stage to improve the detection performance. BotSCL~\cite{wu2023botscl} treats nodes with the same labels as positive samples and other nodes as negative samples. However, these strong setting of hard samples may lead to issues such as model instability and poor generalization.

\begin{figure*}[!t]
\centering
\includegraphics[width=\textwidth]{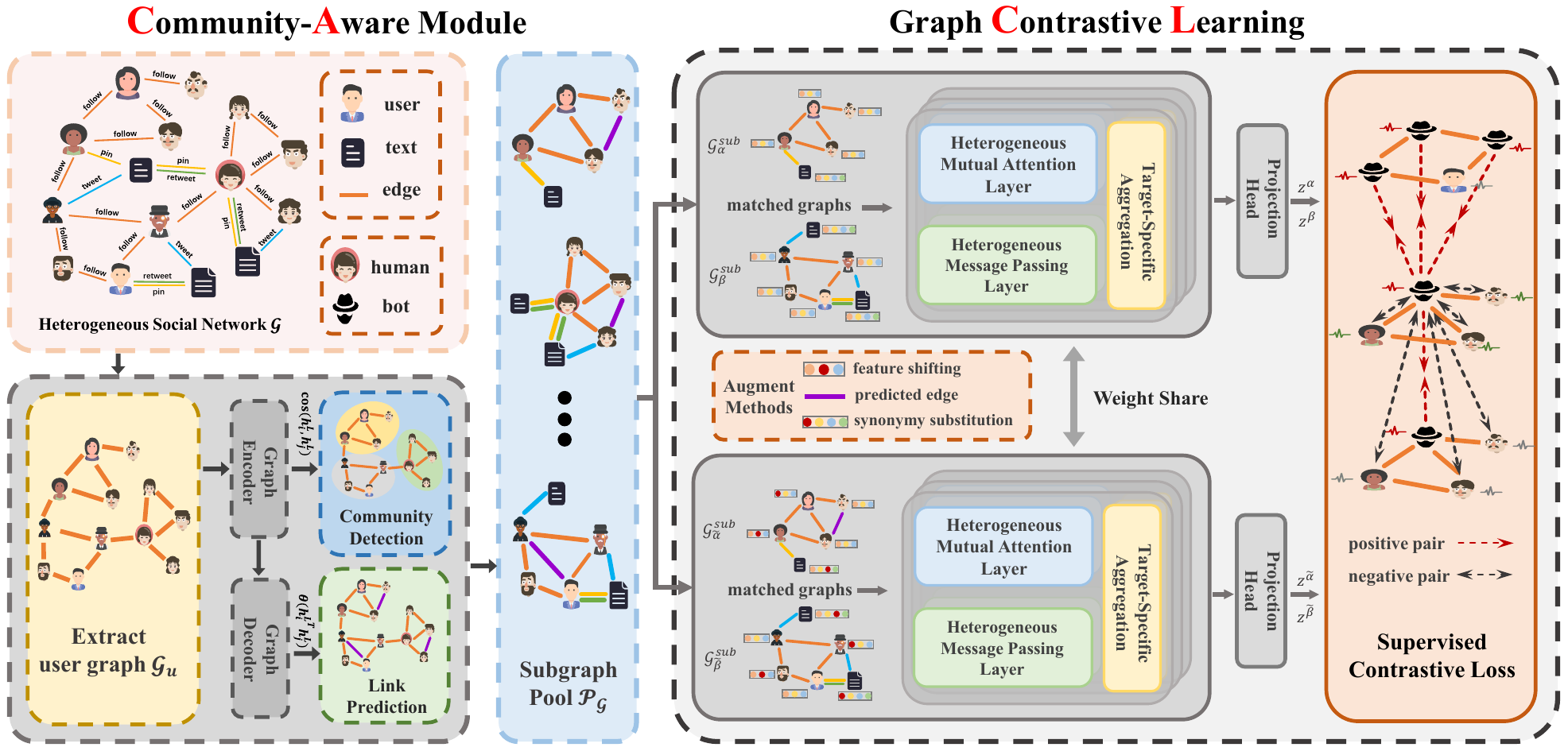}
\caption{\textbf{Overview of our proposed framework CACL}. In graph level, we employ community-aware module to split social networks into communities and mine hard negative and positive samples, creating a subgraph pool. From this pool, we constantly select two matched subgraphs. Then, in node level, we utilize three graph augmentation methods to generate augmented graph and introduce graph contrastive learning to handle hard samples.
}
\label{figure: model_CACL}
\end{figure*}

\section{Methodology}
\subsection{Overview}
\paragraph{Problem Definition.}
We construct a heterogeneous graph $\mathcal{G}=\{\mathcal{V},\mathcal{E},\mathcal{A},\mathcal{R}\}$ where users and texts (e.g., tweets) are treated as nodes $\mathcal{V}=\{v_{i}\mid i=1,2,\ldots,n\}$ with various relationships serving as edges $\mathcal{E}=\{e_{ij}\mid v_i,v_j\in \mathcal{V}\}$. $\mathcal{A}=\{a_i\mid v_i\in \mathcal{V}\}$ collects the types of nodes, and $\mathcal{R}=\{r_{ij}\mid e_{ij}\in \mathcal{E}\}$ collects the types of edges. Social media bot detection task is to learn the distribution of nodes to identify bots among users and apply it to real-world networks.
%Given input $\mathcal{G}_{train}$, social bot detection task is to learn the distribution of nodes to identify bots among users and apply it to real-world networks which are replaced by $\mathcal{G}_{test}$ during the test.

Figure~\ref{figure: model_CACL} displays an overview of our proposed framework \textbf{CACL}. In general, we divide it into coarse-grained and fine-grained operations, corresponding to graph level and node level, respectively. Firstly we perform community detection on the user graph to identify community structures and treat each community as a subgraph. We then perform three graph augmentation methods on subgraphs. Subsequently, we find two matched subgraphs and conduct graph contrastive learning on them to obtain the optimal user representations and finally classify users into bots and humans.

\subsection{Community-Aware Module}
% We follow the traditional definition of community, i.e., the internal degree of each vertex is greater than its external degree\cite{fortunato2010community_define} while ensuring the features of nodes in these subgraphs exhibit high similarity. 
% To capture these community structures in social media networks, we employ an unsupervised graph autoencoder for community detection and link prediction at the same time. Specifically, we utilize the node embedding vectors obtained from the encoder for community detection based on node similarity and graph structure, and then leverage the new edges in graph that reconstructed in the decoder for link prediction. \citep{salha2022modularityCD} have theoretically proved that loss function with modularity-awareness endows graph autoencoder with capability to concurrently accomplish both tasks.

For each input graph $\mathcal{G}$, we first extract user graph $\mathcal{G}_{u}=\{\mathcal{V}_{u},\mathcal{E}_{u},\mathcal{A}_u,\mathcal{R}_u\}$ to simplify the graph structure and generate the adjacency matrix $A$ which disregards different edge types between users. The initial features of users encompass metadata, categorical metadata, descriptions, and posted tweets. We apply z-score normalization to numerical features and use RoBERTa~\cite{liu2019roberta} to encode textual features. We denote user $v_i$'s feature vector as $\mathbf{r}_i$ and preprocess it with a fully connected layer: 
\begin{align}
\mathbf{x}_i=\sigma(\mathbf{W}_{p}\mathbf{r}_i),
\end{align}
where $\mathbf{W}_{p}$ is learnable parameter, and $\sigma$ denotes Leaky-ReLU without further notice.

To incorporate graph structural information into node features and aggregate neighbors' information, we use multi-layer GCNs as encoder to encode the user graph as follows:
\begin{align}
\mathbf{h}^{l+1}=\sigma(A\mathbf{h}^{l}\mathbf{W}^{l}_{h}),
\end{align}
where $\mathbf{h}^{0}_i=\mathbf{x}_i$ is the input of the first layer, $l$ denotes the $l$-th GCNs layer, and $\mathbf{W}^{l}_h$ is learnable parameter. $\mathbf{h}^{L}_i$ is the final representation of $v_i$. 
% During the decode process, for consistency with previous works, we continue using the result of inner product $\hat{A}_{ij}=\theta({\mathbf{x}_i^L}^T\mathbf{x}_j^L)\in[0,1]$ as the probability of an edge existing between user $v_i$ and $v_j$, where $\theta$ is the sigmoid function.

After obtaining users' representations encoded by $L$-layer GCNs, we assign the cosine similarities $cos(\mathbf{h}^L_i,\mathbf{h}^L_j)$ as weights to corresponding edge $e_{ij}$. Then we traverse edges in descending order of weights, consecutively cluster the nodes connected by each edge into same community, and merge overlapped communities until the number of communities equals $k$. To follow the traditional definition of community, i.e., the internal degree of each vertex is greater than its external degree~\cite{fortunato2010community_define}, $k$ is given by Louvain algorithm~\cite{blondel2008louvain} which relies entirely on the inherent graph structure without node features. In this manner, the user graph is partitioned into subgraphs and each subgraph represents a community. Users in the same subgraph have similar representations, and users in different subgraph have different representations, which satisfies the previously definition of hard samples.

To ensure that the community-aware module can simultaneously accomplish community detection task, as well as link prediction used in Section~\ref{sec:graph augmentation}, we refer to~\cite{salha2022modularityCD} and employ a modularity-inspired loss function $\mathcal{L}_{\text{CA}}$ to pre-train community-aware module:
\begin{gather}
{\mathcal{L}}_{\text{CA}} = \mathcal{L}_{\text{G}} + \mathcal{L}_{\text{M}}\label{eq:cd loss},\\
\mathcal{L}_{\text{G}} = -\frac{1}{n^2}\sum_{e_{ij} \in
\mathcal{E}_{u}} \mathbf{CE}(A_{ij}, \hat{A}_{ij}),\\
\mathcal{L}_{\text{M}} = -\frac{\beta}{2m}\sum_{i,j=1}^n\left[A_{ij}-\frac{d_id_j}{2m}\right]e^{-\gamma \Vert \mathbf{h}^{L}_i-\mathbf{h}^{L}_j\Vert^2_2}\label{eq:modularity},
\end{gather}
where $\mathbf{CE}$ denotes cross-entropy function and $d_i$ is the degree of node $v_i$. $\mathcal{L}_{\text{G}}$ is a standard reconstruction loss of graph auto-encoder, while $\mathcal{L}_{\text{M}}$ is a community-aware loss inspired by modularity. In Equation~\ref{eq:modularity}, $n$ denotes the number of users, $m$ denotes the number of edges, and the exponential term, which takes values in the range [0, 1], serves as a softened counterpart of community indicator $\delta_{ij}\in\{0, 1\}$ in original modularity function, where $\delta_{ij}$ equals 1 when $i=j$ and 0 otherwise. The hyperparameter $\beta$ is used to balance the two loss terms and $\gamma$ regulates the magnitude of the exponential term.

In essence, $\mathcal{L}_{\text{M}}$ protects graph structure by ensuring structurally adjacent nodes are more likely to be assigned to one community, while $\mathcal{L}_{\text{G}}$ aims to preserve the performances on link prediction.

\subsection{Graph Augmentation}\label{sec:graph augmentation}
Through community-aware module, we split user graph into a series of user subgraphs. To maintain graph structure, we restore the user subgraphs to original heterogeneous graph and form a subgraph pool $\mathcal{P}_\mathcal{G}=\{\mathcal{G}^{sub}_i\}$. In order to improve the effectiveness of graph contrastive learning, we generate augmented graph $\mathcal{G}^{sub}_{\widetilde{\alpha}}$ after processing orignal graph $\mathcal{G}^{sub}_{\alpha}$ with three methods, including node feature shifting, link prediction, and synonymy substitution.

\paragraph{Node Feature Shifting.} To introduce perturbations on less important node features while preserving crucial ones, we adaptively mask certain dimensions by calculating feature importance using PageRank algorithm~\cite{Page1999pagerank,zhu2021AdaptiveContra}. Formally, we sample a mask vector $\mathbf{m}_a\in \mathbb{R}^{d_a}$ for nodes of node type $a$, where each dimension $\mathbf{m}_o$ is independently drawn from Bernoulli distribution $Bern(1-p_o^f), o=1,2,\dots,d_a$, and $d_a$ is corresponding node feature dimension. We assume that a higher feature value, such as high following number, indicates a greater significance of attribute, which holds true for both discrete and continuous node attributes. So we calculate the weight of dimension $o$ as:
\begin{equation}
    w^f_o=\log\sum_{v_i\in\mathcal{V}} \lvert \mathbf{x}_i \rvert\cdot\rho_c(v_i),
\end{equation}
where $\rho_c(v_i)$ denotes PageRank centrality of node $v_i$. Then we can obtain the probability representing feature importance:
\begin{equation}
    p_o^f = min\left(\frac{w_{max}^f-w_o^f}{w_{max}^f-\overline{w^f}}\cdot p_f,p_\tau\right),
\end{equation}
where hyperparameter $p_f$ controls the magnitude of probability while $p_\tau$ is a truncation value. Finally, we get $v_i$'s augmented feature vector $\widetilde{\mathbf{r}}_i=\mathbf{r}_i\circ \mathbf{m}_{a_i}$, where $\circ$ is Hadamard product.

\paragraph{Link Prediction.} To balance in-degree and out-degree of nodes~\cite{zhou2023linkpred} and enhance the model's robustness against interference from changes in relationships between nodes, we perform link prediction and connect potentially existing edges. For consistency with previous works, we use inner product to act as decoder in the community-aware module and utilize the output as the probability of an edge existing between user $v_i$ and $v_j$, 
\begin{gather}
    \mathcal{E}_\text{pred} = \{e_{ij}\mid \theta({\mathbf{h}_i^L}^T\mathbf{h}_j^L)>p_e\},
\end{gather}
where $\theta$ is the sigmoid function, $p_e$ is hyperparameter used to control the number of edges to prevent irreversible disruption of graph structure caused by excessive edges. Then the augmented edge set $\mathcal{E}_{\widetilde{\alpha}}=\{\mathcal{E}_{\alpha},\mathcal{E}_\text{pred}\}$.

\paragraph{Synonymy Substitution.} Social networks contain a considerable amount of textual information. Through using WordNet~\cite{miller1995wordnet} pre-trained with twitter corpus, we replace words with their synonyms to enhance the texts without altering its fundamental semantics.

\subsection{Graph Convolutional Networks}
Different from homogenous networks, heterogeneous social networks consist of nodes and edges of different types, which requires GNNs to be capable of diverse embedding dimensions and different edge types in information propagation. GNNs methods such as SAGE~\cite{hamilton2017GraphSAGE}, GAT~\cite{velivckovic2017GAT}, HGT~\cite{hu2020HGT} can be employed by our CACL framework. 

We first preprocess user $v_i$'s initial feature $\mathbf{r}_i$ with a fully connected layer:
\begin{gather}
    \hat{\mathbf{x}}_i=\sigma(\mathbf{W}^{a_i}_q\mathbf{r_i}),
\end{gather}
where $\mathbf{W}^{a_i}_q$ is learnable parameter for node type $a_i$. $\mathbf{W}^{u}_q$ for user nodes is shared with $\mathbf{W}_p$ in community-aware module to perform dynamic hard sample mining.
We feed $\hat{\mathbf{x}}_i$ into GNNs model and denote that the $L$-th layer GNNs model outputs the target node hidden state $\mathbf{H}^{L}_t$ which contains information from all its neighbors of different types as well as its own feature in the view of community $\mathcal{G}^{sub}_\alpha$. Therefore, by utilizing these node representations, we can proceed with contrastive learning.

\subsection{Community-Aware Contrastive Loss}
Referring to traditional graph contrastive learning frameworks~\cite{you2020GCL_aug,li2022GCL_idea}, we employs a two-layer perceptron as a projection head to obtain representations for each user $v_i$:
\begin{gather}
    \mathbf{z}_i = \mathbf{W}_2\left(\mathbf{W}_1 \mathbf{H}^{L}_i\right),\label{eq:project}
\end{gather}
where $\mathbf{W}_1$ and $\mathbf{W}_2$ are learnable parameters. For original subgraph $\mathcal{G}^{sub}_\alpha$ and augmented subgraph $\mathcal{G}^{sub}_{\widetilde{\alpha}}$, we can get two representations $\mathbf{z}_i^{\alpha}$ and $\mathbf{z}_i^{\widetilde{\alpha}}$ for the same user node $v_i$. 

To solve the false negatives problems caused by unsupervised contrastive learning~\cite{khosla2020supervised_origin}, \citet{wu2023botscl} leverage annotated information during graph contrastive learning. They try to pull closer all of same-class nodes  and push away all of diff-class nodes in constructed graph.
% Specifically, for node representation $z_i^{1}$ in original graph $\mathcal{G}^{sub}_{1}$, the traditional contrastive loss function is defined as follows:
% \begin{gather}
%     \mathcal{L}^{1}_i = -\frac{1}{N_{y_i}}\sum^{N}_{j=1} \delta_{i,j}\cdot \log \frac{\mathrm{e}^{cos(z^{1}_i,z^{2}_j)/\tau}}{\sum^N_{k=1} \mathrm{e}^{cos(z^{1}_i,z^{2}_k)/\tau}}\nonumber
% \end{gather}
% where $N_{y_i}$ denotes the number of nodes in the same class as node $v_i$, $\tau$ is the temperature coefficient which regulate the degree of distribution uniformity and $\delta_{i,j}$ equals 1 when $v_i$ and $v_j$ belong to same class and 0 otherwise.
In realistic scenarios, however, the differences between different users are often substantial even in the same community. If we unreasonably force high-dimensional semantic features of all different same-class nodes to converge to same point on the hypersphere, it may lead to issues such as model instability and poor generalization. Therefore, we modify the traditional loss function and choose to only push away hard negative samples, which are diff-class nodes in the same community, while pull closer hard positive samples, which are same-class nodes in the different communities just as showing in Figure~\ref{figure:small figure of idea}.

Given one subgraph $\mathcal{G}_{\alpha}^{sub}$, we firstly find its matched subgraph. By computing an average embedding vector for every subgraph and then cosine similarity between each pair of subgraphs in the subgraph pool $\mathcal{P}_{\mathcal{G}}$. Its matched subgraph $\mathcal{G}_{\beta}^{sub}$ has the lowest similarity with $\mathcal{G}_{\alpha}^{sub}$ and thus the positive samples in $\mathcal{G}_{\beta}^{sub}$ is hard enough intuitively. We then consider target node representation $z_i^{\alpha}$ in $\mathcal{G}_{\alpha}^{sub}$, the modified contrastive loss function is as follows:
\begin{gather}
    \mathcal{L}^{\alpha}_i=-\frac{1}{N_{\alpha}}\sum_{v_i\in\mathcal{V}_{\alpha}}\log\frac{s^{self}_i+s^{pos}}{s^{self}_i+s^{pos}_i+s^{neg}_i},\label{eq:loss origin}\\
    \begin{aligned}
        s^{self}_i&=e^{cos(\mathbf{z}^{\alpha}_i,\mathbf{z}^{\widetilde{\alpha}}_i)},\\
        s^{pos}_i&=\frac{1}{2}\sum_{v_j\in\{\mathcal{V}_{\beta},\mathcal{V}_{\widetilde{\beta}}\}}\delta_{ij}e^{cos(\mathbf{z}^{\alpha}_i,\mathbf{z}_j^{\beta,\widetilde{\beta}})},\\
        s^{neg}_i&=\frac{1}{2}\sum_{v_j\in\{\mathcal{V}_{\alpha},\mathcal{V}_{\widetilde{\alpha}}\}}\hat{\delta}_{ij}e^{cos(\mathbf{z}^{\alpha}_i,\mathbf{z}_j^{\alpha,\widetilde{\alpha}})},\\
    \end{aligned}
    \notag
\end{gather}
where $\hat{\delta}_{ij}$ is opposite of ${\delta}_{ij}$. $s^{self}_i$ represents the similarities of target node $v_i$ in different views, $s^{pos}_i$ and $s^{neg}_i$ denote the similarities between target node and its hard positive samples and hard negative samples, respectively. After performing the same operations on the nodes in $\mathcal{G}_{\widetilde{\alpha}}^{sub}$, we obtain the final contrastive loss:
\begin{equation}
    \mathcal{L}_{contrast}=\sum_{\mathcal{G}^{sub}_{\alpha}\in\mathcal{P}_{\mathcal{G}}}\frac{1}{2N_{\alpha}}\sum^{N_{\alpha}}_{i=1} \mathcal{L}_i^{\alpha}+\mathcal{L}_i^{\widetilde{\alpha}},\label{eq:loss contrast}
\end{equation}
where $N_\alpha$ is the number of users in $\mathcal{G}^{sub}_\alpha$. We incorporate this contrastive loss function as a regularization term into the overall model loss function:
\begin{equation}
    \mathcal{L}_{CL} =\lambda\cdot\mathcal{L}_{contrast}+(1-\lambda)\cdot\mathcal{L}_{classify},\label{eq:loss all}
\end{equation}
where the second term represents the cross-entropy classification loss, and $\lambda$ is weight adjustment hyperparameter.

\begin{table*}[t]
    \centering
    
        \resizebox{\linewidth}{!}{
        \begin{tabular}[c]{lccccccccc}
            \toprule[1.5pt]
            \multirow{2}{*}{\textbf{Method}} & \multicolumn{3}{c}{\textbf{Cresci-15}} & \multicolumn{3}{c}{\textbf{Twibot-20}} & \multicolumn{3}{c}{\textbf{Twibot-22}}\\
            \cmidrule[0.75pt](lr){2-4}\cmidrule[0.75pt](lr){5-7}\cmidrule[0.75pt](lr){8-10}
            & \textbf{Accuracy} & \textbf{F1-score} & \textbf{MCC} & \textbf{Accuracy} & \textbf{F1-score} & \textbf{MCC} & \textbf{Accuracy} & \textbf{F1-score} & \textbf{MCC}\\
            \midrule[0.75pt] 
                BotRGCN     & $94.71$ & $95.85$ & $89.06$ & $\underline{84.43}$ & $85.79$ & $67.17$ & $73.90$ & $\underline{48.30}$ & $31.01$\\
                DeeProBot   & $72.35$ & $81.61$ & $44.26$ & $77.34$ & $80.32$ & $54.40$ & $74.11$ & $12.73$ & $14.12$\\
                RGT         & $96.27$ & $\underline{97.04}$ & $\underline{92.25}$ & $84.02$ & $86.07$ & $68.08$ & $74.49$ & $45.36$ & $29.64$\\
                EvolveBot   & $92.18$ & $90.07$ & $72.77$ & $65.83$ & $69.75$ & $30.79$ & $71.09$ & $14.09$ & $13.38$\\
                SimpleHGN   & $93.13$ & $94.68$ & $85.91$ & $83.93$ & $86.33$ & $68.40$ & $72.53$ & $\textbf{49.59}$ & $30.74$\\
                RoBERTa     & $\underline{97.01}$ & $95.86$ & $89.55$ & $75.55$ & $73.09$ & $42.15$ & $72.07$ & $20.53$ & $19.35$\\
                BGSRD       & $87.78$ & $90.80$ & $74.92$ & $66.36$ & $70.05$ & $32.28$ & $71.88$ & $21.14$ & $20.32$\\
                SATAR       & $92.71$ & $94.55$ & $85.61$ & $84.02$ & $85.74$ & $67.83$ & /       & /       & /      \\
                \midrule[0.50pt]
                GAT         & $93.52$ & $94.99$ & $86.78$ & $82.24$ & $84.93$ & $64.95$ & $74.17$ & $43.75$ & $28.12$\\
                +\textbf{CACL}       & $94.51$ & $95.42$ & $88.71$ & $83.52$ & $86.39$ & $68.56$ & $74.50$ & $44.27$ & $28.94$\\
                \midrule[0.50pt]
                SAGE        & $93.92$ & $95.26$ & $87.48$ & $83.00$ & $85.48$ & $66.38$ & $74.39$ & $46.47$ & $30.25$\\
                +\textbf{CACL}       & $\textbf{97.65}$ & $\textbf{98.12}$ & $\textbf{95.10}$ & $83.60$ & $\underline{86.53}$ & $\underline{68.95}$ & $\textbf{75.38}$ & $47.45$ & $\underline{32.27}$\\ 
                \midrule[0.50pt]
                HGT         & $92.24$ & $94.27$ & $84.83$ & $84.02$ & $86.35$ & $68.48$ & $\underline{75.07}$ & $43.13$ & $29.18$\\
                +\textbf{CACL}       & $95.88$ & $96.74$ & $91.45$ & $\textbf{85.12}$ & $\textbf{87.28}$ & $\textbf{70.75}$ & $\underline{75.07}$ & $48.27$ & $\textbf{32.39}$\\
            \bottomrule[1.5pt]
        \end{tabular}
        }
    \caption{\textbf{Main results of CACL compared with various baselines on three datasets}. \textbf{Bold} and \underline{underline} indicate the highest and second highest performance, respectively, while "/" indicates that the baseline is not scalable to the corresponding dataset. GAT, SAGE and HGT are three backbones we use as the graph convolutional model in our framework.}
    \label{tab:main result}
\end{table*}

\subsection{Training Strategies}
In this paper, to mitigate the impact of low-quality community partitions on graph contrastive learning during cold start, we initially perform unsupervised pre-training on community-aware module, using Equation~\ref{eq:cd loss}. In the training process, we iteratively select a certain number of adjacent users to form a input graph. After obtaining the subgraph pool, we continuously retrieve two matched subgraphs from the subgraph pool, and then employ graph contrastive learning for end-to-end model training on these two matched subgraphs using Equation~\ref{eq:loss all}.

\section{Experimental Setups}
\paragraph{Datasets.}
Our method is graph-based and heterogeneous-aware, which requires information of diverse relation types between nodes in datasets. \textbf{Cresci-15}~\cite{cresci2015cresci-15}, \textbf{Twibot-20}~\cite{feng2021twibot20} and \textbf{Twibot-22}~\cite{feng2022twibot-22} all contain various relationships among users as well as between users and their tweets which are essential to build heterogeneous networks. We follow the same splits provided in the benchmarks to mitigate the potential impact on experiment results.

\paragraph{Baselines.}
To demonstrate the effectiveness of our proposed framework \textbf{CACL}, we choose three backbone models, i.e., GCN~\cite{kipf2016GCN}, GAT~\cite{velivckovic2017GAT}, HGT~\cite{hu2020HGT}, as convolution layers and compare them with previous representative works:
\begin{itemize}[leftmargin=*,itemsep=-2pt]
    \item \textbf{SimpleHGN}~\cite{lv2021SimpleHGN} designs residual connections and output embeddings normalization, which is suitable for heterogeneous graphs.
    \item \textbf{BotRGCN}~\cite{feng2021botrgcn} leverages relational graph convolutional networks, enhancing its ability to capture diverse disguises of robots.
    \item \textbf{RGT}~\cite{feng2022RGT} utilizes a multi-attention mechanism and a semantic attention network to learn and aggregate user representations under each relationship, respectively.
    \item \textbf{DeeProBot}~\cite{hayawi2022deeprobot} is designed with LSTM units to process the mixed input, extracting temporal information from user text.
    \item \textbf{EvolveBot}~\cite{yang2013evolvebot} extracts features such as betweenness centrality and clustering coefficient from graph structure.
    \item \textbf{BGSRD}~\cite{guo2021BGSRD} utilizes Roberta to process user descriptions and make prediction with the combination of BERT and GAT outputs.
    \item \textbf{SATAR}~\cite{feng2021satar} considers the follower number as self-supervised label to pre-train models and then fine-tune parameters.
    \item \textbf{RoBERTa}~\cite{liu2019roberta} is a pre-trained language model based on the Transformer architecture, employed by Twibot-22 benchmark.
\end{itemize}
\paragraph{Implementation.} We set $\lambda$ to 0.9, $\tau$ to 0.07, learning rate to 0.0001, layer number of all GNNs models to 2, and dropout rate to 0.5. In each epoch, we iteratively select  2000 adjacent users for Twibot-22 dataset and 1000 for other datasets. We conduct all experiments on 2 NVIDIA RTX A5000 GPUs.
% All of the experiments are conducted in inductive way, which means test set is unseen in training process and it is more in line with the actual situations.

\section{Results and Analysis}

\subsection{Main Results}
We evaluate \textbf{CACL} framework with three backbones and other representative baselines on the three social media bot detection benchmarks and present the results in Table~\ref{tab:main result}. The results demonstrate that: 
\begin{itemize}[leftmargin=*,itemsep=-2pt]
    \item Although previous methods have made great efforts, the improvements of models' performances are not significant compared with traditional GNNs methods.
    \item Our proposed \textbf{CACL} framework markedly enhance the performance of various convolutional networks, including GAT, SAGE and HGT, and can even outperform the state-of-the-art.
    \item \textbf{CACL} framework with different backbones presents significant improvement compared with the original methods, which elucidates that contrastive learning and the community-aware module help GNNs models to better capture node information.
\end{itemize}

\subsection{Contrastive Learning Study}
To compare diverse contrastive learning loss functions, we experiment with three loss functions while keeping other parameters constant. \textbf{Unsupervised contrastive loss}~\cite{you2020GCL_aug} tries to push the representations of all nodes away from each other and pull representations of every node in different views together without using the label information. \textbf{Supervised contrastive loss}~\cite{khosla2020supervised_origin, wu2023botscl} aims to push the representations of diff-class nodes away from each other while pull representations of same-class nodes together, which are conducted across original graph and augmented graph. Our proposed \textbf{CACL} framework utilizes the inherent community structure to mine hard positive and negative samples for bot detection.

\begin{table}[t!]
\centering
\renewcommand\arraystretch{1}
\scalebox{1.}{
\begin{tabular}{lccc}
\hline
\toprule
 \textbf{CL Loss}& \multicolumn{1}{c}{\textbf{Accuracy}} & \multicolumn{1}{c}{\textbf{F1-score}} & \multicolumn{1}{c}{\textbf{MCC}}\\
\midrule[0.75pt] 

unsuper & \multicolumn{1}{c}{84.27} & \multicolumn{1}{c}{86.75} & \multicolumn{1}{c}{69.40}  \\
super    & 84.78$_{\uparrow0.51}$  & 86.95$_{\uparrow0.20}$  & 69.97$_{\uparrow0.57}$  \\
static        & 84.45$_{\uparrow0.18}$  & 86.74$_{\downarrow0.01}$  & 69.42$_{\uparrow0.02}$  \\
dynamic       & 85.12$_{\uparrow0.85}$ & 87.28$_{\uparrow0.53}$ & 70.75$_{\uparrow1.35}$ \\
\bottomrule
\hline
\end{tabular}
}
\caption{\textbf{Performances of HGT model with different contrastive loss function}, including unsupervised and supervised contrastive loss function, whose definitions of hard samples are different from ours, and our modified loss function in \textbf{CACL} framework with static or dynamic community-aware module.}
\label{tab:CL_func}
\end{table}

The comparison results in Table~\ref{tab:CL_func} show that, compared with unsupervised contrastive loss function, other loss functions can significantly improve the performance of HGT model. Our proposed community-aware contrastive loss function outperforms the traditional supervised contrastive loss function, which indicates that mining hard samples under the consideration of community structure is effective for contrastive learning. However, when we freeze community-aware module and thus it remains static during the training process, the improvement decreases significantly, showing the importance of dynamism of this module.

\begin{figure}[t!]
\centering
\resizebox{0.48\textwidth}{!}{
\includegraphics{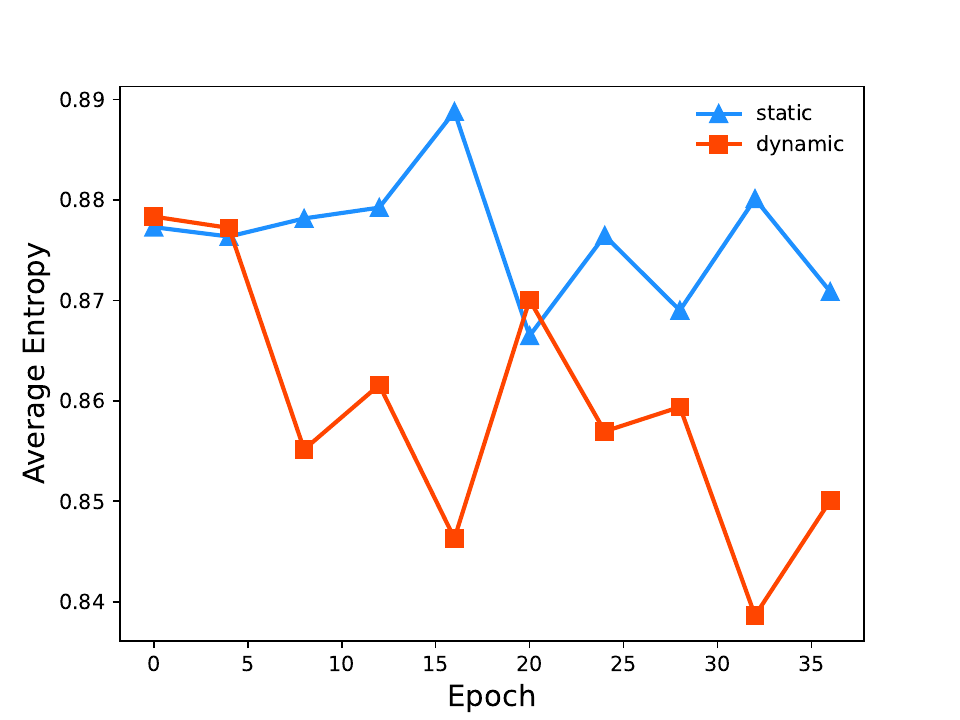}}
\caption{\textbf{Contrast of community entropy change trends} between static and dynamic community-aware module. Static means freezing community-aware module and thus cannot dynamically mine hard samples during training process. Average entropy is a metric for measuring the imbalance of categories.
}
\label{figure: entropy_contrast}
\end{figure}

\subsection{Community-aware Study}
To investigate the internal mechanism of how community-aware module helps model to focus on hard samples and classify them correctly, we use information entropy, which is adopted in decision trees~\cite{quinlan1986ID3}, to measure the level of category disorder in each community we detect in every batch. To demonstrate the overall trend, we take average of entropy in each epoch and make contrast between dynamic and static community-aware module. The results shown in Figure~\ref{figure: entropy_contrast} demonstrate that dynamic community-aware module tend to cluster nodes of same labels to the same community, which proves it can dynamically capture the community structures within social networks as parameters are continuously updated in the training process, while the static module fails.

\begin{figure}[t!]
\centering
\resizebox{0.475\textwidth}{!}{
\includegraphics{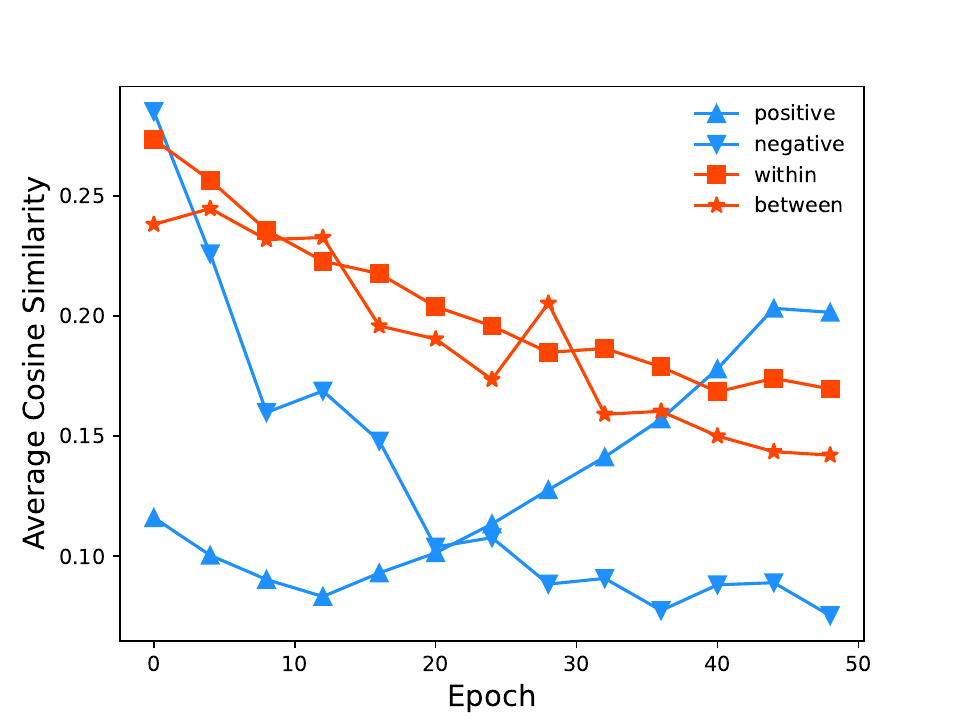}}
\caption{\textbf{Cosine similarity change trends}. Positive and negative represent cosine similarities between positive samples and between negative samples, respectively. Within and between represent cosine similarities within one community and between different communities, respectively. 
}
\label{figure: cos_change_contrast}
\end{figure}

To further qualitatively analyze whether the community detection result is meaningful, and how the contrastive learning design helps to deal with the hard samples, we visualize the changes of average cosine similarity among nodes in one community and nodes in different communities. The results shown in Figure~\ref{figure: cos_change_contrast} prove that the similarities between positive samples continuously rise and those between negative samples decrease, which means same-class nodes are pulled together and diff-class nodes are pushed apart. The results also show that similarities of nodes within and between communities decrease as we dynamically update the community detection module, which proves the community detection can dynamically capture hard samples and the result is helpful for the detection.

\begin{table}[t!]
\centering
\renewcommand\arraystretch{1}
\scalebox{1.}{
\begin{tabular}{lccc}
\hline
\toprule
 \textbf{Ablation}& \multicolumn{1}{c}{\textbf{Accuracy}} & \multicolumn{1}{c}{\textbf{F1-score}} & \multicolumn{1}{c}{\textbf{MCC}}\\
\midrule[0.75pt] 
\textbf{CACL} & \multicolumn{1}{c}{85.12} & \multicolumn{1}{c}{87.28} & \multicolumn{1}{c}{70.75}  \\
\midrule[0.50pt]
w/o SS    & 84.78$_{\downarrow0.34}$  & 87.19$_{\downarrow0.09}$  & 87.19$_{\downarrow0.24}$  \\
w/o LP        & 84.45$_{\downarrow0.67}$  & 86.98$_{\downarrow0.30}$  & 70.01$_{\downarrow0.74}$  \\
w/o FS       & 85.12$_{\downarrow0.00}$ & 87.09$_{\downarrow0.19}$ & 70.42$_{\downarrow0.33}$ \\
\midrule[0.50pt]
w/o TX     & 83.34$_{\downarrow1.78}$ & 86.51$_{\downarrow0.77}$ & 69.14$_{\downarrow1.61}$ \\
w/o MT     & 82.92$_{\downarrow2.20}$ & 86.10$_{\downarrow1.18}$ & 67.97$_{\downarrow2.78}$ \\
w/o HG     & 82.75$_{\downarrow2.37}$ & 85.28$_{\downarrow2.00}$ & 65.87$_{\downarrow4.88}$ \\
\midrule[0.50pt]
w/o CA     & 84.36$_{\downarrow0.76}$ & 86.44$_{\downarrow0.84}$ & 68.87$_{\downarrow1.88}$ \\
\bottomrule
\hline
\end{tabular}
}
\caption{
\textbf{Ablation study of CACL.}
Experiments mainly focus on there aspects. SS, LP, FS symbolizes three data augmentation methods, which are synonymy substitution, link prediction and feature shifting, respectively. 
TX, MT, HG indicates three different kinds of information in social networks, which are users' texts, metadata and heterogeneity of graph.
CA denotes the community-aware hard samples mining.
}
\label{tab:ablation}
\end{table}

\subsection{Ablation Study}
As \textbf{CACL} framework can improve the performances of all three backbones, we conduct ablation study to analysis the effects of different model components as well as data structure. We use HGT as backbone and present the results in Table~\ref{tab:ablation}:
\begin{itemize}[leftmargin=*,itemsep=-2pt]
    \item To study the impact graph augmentation methods, we remove each method individually. Masking node feature shifting results in the least decrease in accuracy compared to other methods. Removing link prediction leads to a sharp decrease in performance, indicating the importance of finding potential social relationships.
    \item To evaluate the effectiveness of leveraging multiple user information modalities, we mask users' metadata, texts and remove the graph's heterogeneity separately. Compared with the full model, any absence of information from the social networks results in decreased performance, which reveals the importance of all these kinds of user information for bot detection.
    \item To investigate the importance of community-aware hard samples mining methods, we mask the community-aware module and process the origin graph with GNNs directly. Final results show a decrease in performance, which proves the significance of community-aware module.
\end{itemize}

\section{Conclusion}
Social media bot detection is an application problem that integrates various fields. To leverage the rich community structure information in social networks and address over-smoothness problem in GNNs model, we propose the \textbf{CACL} framework, which utilizes community-aware module to mine hard positive and negative samples at graph level, and introduces graph contrastive learning to handle hard samples in node level. \textbf{CACL} also employs three efficient graph augmentation methods which enhance GNNs model's robustness and generalization ability. We conducte extensive experiments on three comprehensive benchmarks which demonstrate the effectiveness of our framework.

\section*{Limitations}
In this paper, we mainly focus on constructing a community-aware graph contrastive learning framework. Despite our best efforts, this paper may still have some limitations. We evaluate our method in three benchmarks, including the most comprehensive dataset Twibot-22 in Twitter bot detection area. However, \citet{hays2023bot_det_prob} still point out that current datasets still have certain limitations in collecting and labeling that affect the evaluation of methods. If we find several communities in real world and record the changes of them during the training process, we can simply reveal the effectiveness of our proposed framework. However, due to our framework is dynamic updated, community structure is dynamic and hard to record.

Large language models exhibit a high level of comprehension ability for graph information. Effectively integrating LLM technology can enhance the extraction of user information, which contributes to the identification of robot identities. For example, \citet{liu2023botmoe} use mixture of experts to improve the performance and \citet{tang2023graphgpt} employ GPT to deal with graph data. Despite their excellent performances, we focus on enhancing the performance of GNNs models, so we do not delve into these aspects too much in this paper.

\section*{Ethics Statement}
Our research in this paper follows ethical guidelines and principles. All experiments and data usage are conducted in accordance with the relevant institutional and national guidelines and regulations. Despite the datasets we use contain a large amount of private information existing in social media, we ensure the privacy and confidentiality of any personal or sensitive information used and obtained during the study. Our work aims to contribute positively to online environment, while respecting the dignity, rights, and well-being of individuals and communities. We acknowledge and address any potential biases or ethical considerations in our research.

\section*{Acknowledgements}
This research is supported by the National Natural Science Foundation of China (No.62106105), the CCF-Baidu Open Fund (No.CCF-Baidu202307), the CCF-Zhipu AI Large Model Fund (No.CCF-Zhipu202315), the Scientific Research Starting Foundation of Nanjing University of Aeronautics and Astronautics (No.YQR21022),  the High Performance Computing Platform of Nanjing University of Aeronautics and Astronautics, and the foundation of National College Students' innovation and entrepreneurship training program (No.202310287085Z).

\bibliography{custom}
\bibliographystyle{acl_natbib}

\end{document}